\documentclass[11pt,a4paper]{article} 
\usepackage{amsmath}     
\usepackage{amssymb}
\usepackage{amsthm}
\usepackage{eurosym}
\usepackage{setspace}
\usepackage{url}
\usepackage{doi}

\DeclareMathOperator*{\argmax}{argmax} 

\usepackage[super,comma,sort&compress]{natbib}
\usepackage{sectsty}           
    \sectionfont{\large}       

\usepackage{xcolor}

\usepackage{sgame, tikz} 
\usetikzlibrary{trees, calc, shapes} 
\usepackage{amsfonts, mathtools}

\usepackage{caption}
\usepackage{float}
\usepackage{subfig}             
\newcommand\T{\rule{0pt}{2.6ex}}
\newcommand\B{\rule[-1.2ex]{0pt}{0pt}}
\newcommand\BB{\rule[-2.4ex]{0pt}{0pt}}

       \usepackage[a4paper]{anysize} 
       \marginsize{2.5cm}{2.5cm}{2.5cm}{.8cm}

\numberwithin{equation}{section}  

\theoremstyle{plain}

\theoremstyle{definition}

\theoremstyle{remark}

\hypersetup{
	colorlinks,
	linkcolor={red!50!black},
	citecolor={blue!50!black},
	urlcolor={blue!80!black}
}

\begin{document}
\linespread{1.3}\small\normalsize
\title{\textsc{\Large{
Many Phish in the $\mathcal{C}$\,:\\
A Coexisting-Choice-Criteria Model of Security Behavior%
 \thanks{Copyright \copyright{} 2018 Iain Embrey and Kim Kaivanto }
}}}

 \author{Iain Embrey%
\thanks{Financial support from the Economic and Social Research Council (UK) made this work possible, and is gratefully acknowledged (grant number ES/P000665/1.).}
~and Kim Kaivanto\\
 Department of Economics, Lancaster University, Lancaster LA1 4YX, UK
}

\date{\emph{this version:} \today}

\maketitle
\thispagestyle{empty} 

\begin{center}{\bf Abstract}\end{center} 
Normative decision theory proves inadequate for modeling human responses to the social-engineering campaigns of Advanced Persistent Threat (APT) attacks.
Behavioral decision theory fares better, but still falls short of capturing social-engineering attack vectors, which operate through emotions and peripheral-route persuasion.
We introduce a generalized decision theory, under which any decision will be made according to one of multiple coexisting choice criteria.
We denote the set of possible choice criteria by $\mathcal{C}$. 
Thus the proposed model reduces to conventional Expected Utility theory when $|\,\mathcal{C}_{\text{EU}}|=1$, whilst Dual-Process (thinking fast vs. thinking slow) decision making corresponds to a model with $|\,\mathcal{C}_{\text{DP}}|=2$. We consider a more general case with $|\,\mathcal{C}|\geq 2$, which necessitates careful consideration of \emph{how}, for a particular choice-task instance, one criterion comes to prevail over others.
We operationalize this with a probability distribution that is conditional upon traits of the decision maker as well as upon the context and the framing of choice options. 
Whereas existing Signal Detection Theory (SDT) models of phishing detection commingle the different peripheral-route persuasion pathways, in the present descriptive generalization the different pathways are explicitly identified and represented. 
A number of implications follow immediately from this formulation, ranging from the conditional nature of security-breach risk to delineation of the prerequisites for valid tests of security training.
Moreover, the model explains the `stepping-stone' penetration pattern of APT attacks, which  has confounded modeling approaches based on normative rationality. 

\begin{spacing}{1.5}

\end{spacing}

\vfill
\smallskip
\noindent \emph{Keywords:}\/\;
phishing,
social engineering,
peripheral-route persuasion,
advanced persistent threat,
choice criteria,
dual-process theory,
latent class model

\noindent \emph{JEL classification:}\/\;
D81,
D91

\medskip

\pagebreak
\doublespacing
\section{INTRODUCTION}
The human element in decision making is not only deliberative, but also emotional, intuitive, and fallible.
Social-engineering campaigns target and exploit these non-deliberative features of human decision making.$^{(}$\!\citep{petty/cacioppo:86,rusch:99,langenderfer/shimp:01-pm,mitnick/simon:02,cialdini:07,hadnagy:11,oliveira/etal:17}$^{\!)}$
A major lacuna for security-behavior modeling is that standard decision theory fails to capture the peripheral-route persuasion pathways 
that are exploited in social-engineering campaigns.

In contrast, Signal Detection Theory (SDT) has been successfully adapted to model human responses to phishing attacks.\footnote{In a phishing attack, a network user receives an email containing either an attachment or a website link, which if opened, prompts the user to enter personal information (e.g. passwords) or infects the user's computer with malware that records such information surreptitiously. For SDT-based models of phishing vulnerability, see Kaivanto,$^{(}$\!\citep{kaivanto:14-ra}$^{\!)}$ and Canfield and Fischhoff.$^{(}$\!\citep{canfield/fischhoff:17-ra}$^{\!)}$}
The flexibility of SDT is instrumental in this context.
This flexibility has been exploited to study the distinct consequences for security-breach risk estimates of premising the model
solely upon normative decision theory, solely upon behavioral decision theory, or upon the combination of behavioral decision theory and susceptibility to peripheral-route persuasion.$^{(}$\!\citep{kaivanto:14-ra}$^{\!)}$
Unsurprisingly, the latter combination proves most useful and informative.
Nevertheless, two limitations may be observed in the existing SDT-based approach:
(i) decision makers are assumed to be permanently characterized by one fixed decision-making model, and
(ii) the effects of different peripheral-route persuasion pathways feed into, and become commingled in, a single value of the discriminability parameter.\footnote{The standardized distance between the means of the sampling distributions, respectfully under the null and alternative hypotheses.}
Descriptive validity favors relaxation of the former, while interpretability of modeling favors relaxation of the latter.

We introduce a generalization of decision theory that fulfills these desiderata.%
\footnote{The theory we present here is a specialization of Iain Embrey's `States of Mind and States of Nature' formulation.$^{(}$\!\citep{embrey:17}$^{\!)}$}
 %
 %
The generalization comprises two principal components.

First, a non-degenerate set $\mathcal{C}$ of `ways of deciding' -- here called `choice criteria' -- which in the phishing context includes
not only Expected Utility (EU) to capture rational deliberative decision making,
but also Prospect Theory (PT) which captures behavioral decision-making,%
$^{(}$\!\citep{tversky/kahneman:92-jru}$^{\!)}$
a `routinely click-straight-through' element that captures unmotivated and unthinking routinized actions (automaticity),%
$^{(}$\!\citep{moors/dehouwer:06-pb}$^{\!)}$
and an `impulsively click-through' element that captures emotionally motivated impulsive actions.%
$^{(}$\!\citep{petty/cacioppo:86,rusch:99,langenderfer/shimp:01-pm,mitnick/simon:02,cialdini:07,hadnagy:11,oliveira/etal:17}$^{\!)}$
This approach therefore generalizes not only EU and PT, but also Dual-Process (DP) theories.\footnote{$\mathcal{C}_{\text{Here}}\supset\mathcal{C}_{\text{DP}}\supset \mathcal{C}_{\text{EU}},\,\mathcal{C}_{\text{PT}}$.} 

It also formalizes the notion -- to which the paper's title alludes -- that there are several distinct types or classes of phishing ploy, and that individuals' susceptibility differs across qualitatively distinct social-engineering attack vectors. 
It is important to distinguish between these distinct phishing attack vectors -- both to understand individuals' behavioral responses to them, and to understand organizations' total security-breach risk exposure.
A phishing ploy that plays upon the prospect of a time-delimited opportunity for wealth is constructed very differently -- and is processed very differently by its recipient(s) -- than a phishing ploy that plays upon employees' standard routines of unquestioningly responding to bosses' and colleagues' emails, opening any appended email attachments, and clicking on enclosed links.
An organization's email security training may effectively address the former, but in many organizations the latter remains a worrying vulnerability.

The second component of the generalization is a conditional probability distribution over the different choice criteria, i.e. over the elements of the set 
$\mathcal{C}$. 
As each new choice task is confronted, a draw from this distribution determines which choice criterion becomes operative, and so we will refer to it as the \emph{State-of-Mind} (SoM) distribution for an individual $i$ at time $t$.  
We allow an individual's SoM distribution to be conditional upon: their  psychological traits and decision-experiences, the situational context of the decision, and the framing of the choice options. 
This approach is similar to that of two existing addiction models$^{(}$\!\citep{laibson:01-qje,bernheim/rangel:04-aer}$^{\!)}$ although we extend those models by allowing the framing of the choice options to be strategically determined by an adversarial agent (the attacker), and by allowing both the prior-experiences and situational context of a decision to be strategically influenced by an allied agent (the Information Security Officer).

A key advantage of the present formulation is the top-level differentiation of the decision maker's susceptibility to different kinds of phishing ploys.
This formulation yields a number of immediate implications.
First, the overall security-breach risk due to phishing can not be conceived in unconditional terms.
Since an individual's susceptibility to phishing depends on the type of phishing ploy, the phishing-ploy-\emph{type} exposure distribution takes on importance, as does the intensity of this exposure (i.e. the total number of phishing emails traversing the spam filter) and the quality of phishing-ploy execution. 
Second, a single test-phishing email is insufficient for evaluating the effectiveness of email security training. 
Email security training does not necessarily generalize across different choice criteria.
Hence, a single test-phishing email may determine the robustness of security practice towards one particular phishing ploy, but it is orthogonal to potential vulnerabilities within the remaining choice criteria. 
Third, not only is the organization's security-breach risk conditional, but the attacker gets to \emph{choose} the  phishing-ploy-type exposure distribution, as well as the intensity of this exposure. The attacker has first-mover advantage.  
Moreover, the attacker always has the option to develop \emph{new} phishing-ploy types that are not addressed by the organization's existing working practices and training materials. 
Fourth, given working practices in most organizations and given the dimensions over which the attacker can tailor a phishing campaign, it is clear that the attacker can attain a very high total probability of successfully breaching the target organization's cybersecurity. 
In part, this is due to the fact that typical working practices in non-high-security organizations%
\footnote{e.g. commercial, administrative, professional-service, and higher-education organizations} 
do not involve special treatment of embedded links or attached files.%
\footnote{For instance the production of research articles involves multiple exchanges of emails among coauthors themselves, between the coauthors and the journal's editorial team, and then between the coauthors and the publisher's production team. These emails contain file attachments, and sometimes URLs as well. In these exchanges, there is no security procedure in place to authenticate emails, their attachments, or embedded URLs.}
It is also due to the \emph{disjunctive} accumulation (addition, rather than multiplication) of successful-security-breach probabilities over spam-filter-traversing phishing emails. 
But it is also due to the scope for using rich contextual information to tailor a campaign into a \emph{spear-phishing} attack -- i.e. to specifically target the `routinely click-straight-through' choice criterion characterized by automaticity. 

Furthermore, our model supports an explanation for the `stepping-stone penetration pattern' that is common in APT attacks.
Whereas models of security behavior premised upon normative rationality have not been successful in explaining the stepping-stone pattern, we show that in light of a coexisting-choice-criteria model of security behavior, the stepping-stone penetration pattern may be recovered as a constrained-optimal attack vector. 

The sequel is organized as follows.
Section \ref{sec:human-element} briefly reviews the phishing literature, showing that phishing attacks employ social-engineering techniques that circumvent deliberatively rational decision processes. 
Section \ref{sec:ways-of-deciding-provenance} reviews the empirical literature in which multiple `ways of deciding' have been documented empirically, establishing a rigorous empirically grounded basis for the coexisting-choice-criteria model. 
Section \ref{sec:homo-intuivus} introduces the coexisting-choice-criteria model, and illustrates some of its properties, including its ability to support an explanation of the stepping-stone penetration pattern (Section \ref{sec:stepping-stone-penetration}). 
Section 5 concludes. 

\section{PHISHING TARGETS THE HUMAN ELEMENT}
\label{sec:human-element}
The capacity for rational deliberation is a feature of human beings, albeit not the overriding trait it was thought to be when Carl Linnaeus coined the binary nomenclature, \emph{Homo sapiens}.\footnote{Sapience denotes wisdom and the capacity for sound judgment, particularly in complex or difficult circumstances.}
Both large-scale and narrowly targeted social engineering are predicated upon the intuitive, emotional, and fallible nature of human behavior, and it is now recognized that psychology is an essential component -- alongside engineering and economics -- for understanding information security.$^{(}$\!\citep{anderson/moore:09-ptrsa}$^{\!)}$ 

More than half of all US government network security-incident reports concern phishing attacks, and the number of phishing emails being sent to users of federal networks is growing rapidly.$^{(}$\!\citep{usomb2012,johnson:13-fedtim}$^{\!)}$
The FBI and the DHS recently issued an amber alert warning of APT activity targeting energy -- especially, nuclear power\footnote{For instance, the non-operational computer systems of Wolf Creek Nuclear Operating Corporation in Kansas were penetrated.$^{(}$\!\citep{perlroth:17-nyt}$^{\!)}$} -- and other sectors.$^{(}$\!\citep{fbi/dhs:17}$^{\!)}$
In this broad APT campaign, spear phishing was the preferred initial-breach technique. 
The corporate sector is targeted more widely, commonly using phishing to create an entry point, for the purposes of extortion, illegally acquiring customer-information (and credentials) databases, as well as for acquiring commercially sensitive information. 
The incidence of corporate cyber espionage is not systematically disclosed, but many of the high-profile examples of corporate hacking that have come into the public domain were staged via phishing.$^{(}$\!\citep{elgin/lawrence/riley:12-b}$^{\!)}$

Online scams such as phishing and spear phishing employ techniques of persuasion that have collectively been labeled `social engineering'.$^{(}$\!\citep{rusch:99,hadnagy:11}$^{\!)}$
These techniques eschew direct, rational argumentation in favor of `peripheral' routes to persuasion.
The most prominent of these peripheral pathways to persuasion are, in no particular order:
(i) authority,
(ii) scarcity,
(iii) similarity and identification,
(iv) reciprocation,
(v) consistency following commitment, and
(vi) social proof.$^{(}$\!\citep{petty/cacioppo:86,rusch:99,langenderfer/shimp:01-pm,mitnick/simon:02,cialdini:07,hadnagy:11,oliveira/etal:17}$^{\!)}$
Scams\footnote{as well as `hard-sell' and `high-pressure' marketing more generally,} typically augment peripheral-route persuasion by setting up a scenario that creates psychological pressure by triggering \emph{visceral emotions} that override rational deliberation.$^{(}$\!\citep{loewenstein:96-obhp,loewenstein:00-aer,langenderfer/shimp:01-pm}$^{\!)}$
Visceral emotions -- such as greed, envy, pity, lust, fear and anxiety -- generate psychological discomfort as long as the underlying need remains unfulfilled, and psychological pleasure or even euphoria when that need is fulfilled.
The manipulative scenario is deliberately structured so that the scammer's proposition offers the double prospect of relief from the visceral discomfort as well as visceral satisfaction upon fulfilling the underlying need.

An ideally scripted scam scenario contrives a compelling, credible need for immediate action.
If a scam-scenario script falls short of this ideal, it will almost invariably emphasize the urgency with which action must be taken.$^{(}$\!\citep{langenderfer/shimp:01-pm,loewenstein:96-obhp,loewenstein:00-aer}$^{\!)}$
In itself, this introduces visceral anxiety where none existed before, and simultaneously, precludes the availability of time for cooling off and for rational deliberation.
Visceral emotions have both a direct hedonic impact as well as an impact via altering the relative desirability of different cues and attributes.
Crucially, visceral emotions also affect how decision makers process information, narrowing and restricting attention to the focal hedonic cue and its availability (or absence) in the present.$^{(}$\!\citep{loewenstein:96-obhp,loewenstein:00-aer}$^{\!)}$
Since visceral emotions -- and their concomitant effects on attention and relative desirability of different cues/attributes -- are short lived, scam scripts contrive reasons for immediate action.\footnote{A former swindler relates the principle: ``It is imperative that you work as quickly as possible. Never give a hot mooch time to cool off. You want to close him while he is still slobbering with greed.''$^{(}$\!\citep{easley:94}$^{\!)}$}

At sufficiently high levels of intensity, visceral emotions can override rational deliberation entirely.$^{(}$\!\citep{loewenstein:96-obhp}$^{\!)}$
Mass phishing scams often aim to exploit human emotions in this fashion.
Spear phishing attacks, on the other hand, typically aim to exploit the intuitive and fallible nature of human decision making without necessarily stoking emotion.
This approach targets the routinization and \emph{automaticity}$^{(}$\!\citep{moors/dehouwer:06-pb}$^{\!)}$ upon which successful management of a high-volume inbox rests.
For most civilian organizations outside the security community, employees trust emails -- and any embedded URLs and file attachments -- sent by bosses and immediate colleagues, and frequently also those sent by more distant contacts. Failure to do so would bring most organizations to a halt.
Spear phishing thus exploits this routine and unquestioning trust that is automatically extended to bosses, colleagues, and contacts -- and unintendedly, to plausible facsimiles thereof.

More surprising is the fact that spear phishing emails endowed with rich contextual information have been deployed successfully on both sides of the civilian/non-civilian and security/non-security divides.
A partial list of successfully breached governmental, defense, corporate, and scientific organizations includes the White House, the Australian Government, the Reserve Bank of Australia, the Canadian Government, the Epsilon mailing list service, Gmail, Lockheed Martin, Oak Ridge National Laboratory, RSA SecureID, Coka Cola Co., Chesapeake Energy, and Wolf Creek Nuclear Operating Corporation.$^{(}$\!\citep{usomb2012,johnson:13-fedtim,elgin/lawrence/riley:12-b,hong:12-cacm,perlroth:17-nyt}$^{\!)}$
When implemented well with appropriate contextual information, a spear-phishing email simply does not attract critical evaluation, and its contents are acted upon in a routine and automatic fashion.

\section{COEXISTING CHOICE CRITERIA: EMPIRICAL PROVENANCE}
\label{sec:ways-of-deciding-provenance}
Decision theorists are gradually coming to terms with the implications of dual-process theory, which has been developed by psychologists and recently popularized by Daniel Kahneman in \emph{Thinking, Fast and Slow}.$^{(}$\!\citep{kahneman:12}$^{\!)}$

Meanwhile, a well-established stream of empirical-decision-theory literature offers legitimation for the notion that there may be more than one way of reaching a decision. That literature captures heterogeneity in choice criteria with Finite Mixture (FM) models.
Standard estimation procedures for such models allow the data to determine how many different choice criteria are present, and then to provide, for each individual, the respective criterion-type membership probabilities.
In Glenn Harrison and Elisabet Rutstr\"{o}m's FM models,\footnote{of decision making under risk} the traditional single-criterion specification is statistically rejected, in their words providing ``a decent funeral for the representative agent model that assumes only one type of decision process.''$^{(}$\!\citep{harrison/rutstrom:09-ee}$^{\!)}$
In turn, Coller et al.'s FM models show that ``observed choices in discounting experiments are consistent with roughly one-half of the subjects using exponential discounting and one-half using quasi-hyperbolic discounting.''$^{(}$\!\citep{coller/etal:11-oep}$^{\!)}$
And using a Bayesian approach, Houser et al. show that different people use different decision rules -- specifically, one of three different criteria -- when solving dynamic decision problems.$^{(}$\!\citep{houser/etal:04-ec}$^{\!)}$

Multiple choice criteria are also well established in the empirical-game-theory literature.
Stahl and Wilson fit an FM model to data on play in several classes of of 3$\times$3 normal-form games, and find that players fall into five different boundedly rational choice-criteria classes.$^{(}$\!\citep{stahl/wilson:95-geb}$^{\!)}$
Guessing games -- a.k.a. Beauty-Contest games -- have been pivotal in showing not only that backward induction and dominance-solvability break down, but also that game play can be characterized by membership in a boundedly rational, discrete (level-$k$) depth-of-reasoning class.$^{(}$\!\citep{nagel:95-aer}$^{\!)}$
FM models are the technique of choice for analyzing Beauty-Contest data, revealing that virtually all `non-theorist' subjects\footnote{those who are not professional game theorists} (94\%) fall into one of three boundedly rational depth-of-reasoning classes (levels 0, 1 or 2).$^{(}$\!\citep{stahl:96-geb,bosch-domenech/etal:10-ee}$^{\!)}$
FM models are being applied increasingly in empirical game theory -- including to the analysis of e.g. trust-game data, social-preferences data, and common-pool-resource data -- demonstrating the broad applicability of a multiple-criteria approach. 
The theoretical relevance of level-\emph{k} reasoning to adversarial interactions such as phishing has been further demonstrated by Rothschild et al.,$^{(}$\!\citep{rothschild/etal:12-ra}$^{\!)}$ however we know of no existing paper in this field that allows alternative choice criteria to coexist.

Outside decision theory and empirical game theory, the necessity of allowing for multiple choice criteria has also been recognized in the fields of transportation research and consumer research.
Within a Latent Class (LC) model framework,\footnote{Latent Class (LC) models are  specializations of FM models.} Hess et al. study the question of whether ``actual behavioral processes used in making a choice may in fact vary across respondents within a single dataset.''$^{(}$\!\citep{hess/etal:12-t}$^{\!)}$
Preference heterogeneity documented in conventional single-choice-criterion models\footnote{e.g. heterogeneity in risk aversion in EU models, and heterogeneity in probability weighting in PT models} may be a logical consequence of the single-choice-criterion restriction (i.e. misspecification).
Hess et al. account for choice-criterion heterogeneity in four different transport-mode-choice datasets by fitting LC models.
These LC models distinguish
between conventional random utility and the lexicographic choice criterion (dataset 1),
among choice criteria with different reference points (dataset 2),\footnote{note that standard random utility has no reference point}
between standard random utility and the elimination-by-aspects choice criterion (dataset 3),
and between standard random utility and the random-regret choice criterion (dataset 4).$^{(}$\!\citep{hess/etal:12-t}$^{\!)}$
Finally, Swait and Adamowicz show that \emph{consumers} also fall into different `decision strategy' LCs, and that increasing either the complexity of the choice task or the cumulative task burden induces switching toward simpler decision strategies.$^{(}$\!\citep{swait/adamowicz:01-jcr}$^{\!)}$
These results underscore an interpretation of the choice-criterion probabilities that is only implicit in the above-mentioned studies: that
(a) decision makers should not be characterized solely in terms of their \emph{modal} choice criterion, but in terms of their choice-criterion mixtures, and that
(b) the criterion that is operative for a particular choice task is obtained as a draw from the probability distribution over choice criteria, which in turn is conditional upon features of the context, the framing and presentation of the choice options,
and the current psychological characteristics of the decision maker.

In light of these FM- and LC-model findings, accommodation of multiple choice criteria emerges as a natural step toward improving the descriptive validity of theoretical models.

\section{INCORPORATING \emph{Homo intuivus, emotus et fallibilis}}
\label{sec:homo-intuivus}
\subsection{Coexisting-choice-criteria model}
\label{sec:cccm}
The econometric evidence reviewed in Section \ref{sec:ways-of-deciding-provenance} warrants a generalization of decision theory to incorporate multiple coexisting choice criteria. 
An abstract formulation of such a theory naturally draws upon the formal specification of econometric latent class models that capture choice-criterion heterogeneity.$^{(}$\!\citep{hess/etal:12-t,swait/adamowicz:01-jcr}$^{\!)}$ 

Let  $\mathcal{C}$ denote the set of coexisting choice criteria. 
The elements of this set are distinguished by the integer-valued index $c$, where  $1 \leq c \leq C := |\,\mathcal{C}|$.  

We specialize the present formulation to the context of phishing-security modeling by populating the set of choice criteria $\mathcal{C}$ with a view to capturing the essential features of human beings in the security setting, as reviewed in Section \ref{sec:human-element}. 
Email recipients are capable of rational deliberation, but they are not overwhelmingly predisposed to it. 
They may instead form subjective beliefs and valuations as captured by behavioral decision theory, but they also frequently act in an intuitive or routinized fashion. 
Thus the empirical evidence reviewed in Section \ref{sec:human-element} suggests that human responses to phishing campaigns range across (at least) four identifiable choice criteria, which we summarize in Table \ref{tab:ccc}.\footnote{Although the present formulation moves beyond the simplicity of a single-decision-criterion world view, each criterion of Table \ref{tab:ccc} can be formalized by an existing theoretical framework. 
Normative deliberation is underpinned by the axiomatizations of e.g. von Neumann and Morgenstern, or Leonard Savage. 
Descriptively valid, partly deliberative behavioral rationality is underpinned by axiomatizations of Cumulative Prospect Theory by e.g. Wakker and Tversky.$^{(}$\!\citep{wakker/tversky:93-jru}$^{\!)}$
The deliberative-rationality-displacing role of visceral emotions has been recognized in the evolutionary study of behavior, represented in economics in particular by e.g. Robert H. Frank.$^{(}$\!\citep{frank:88}$^{\!)}$
Automaticity, in which deliberative rationality is not so much bypassed as simply `not engaged', has been given theoretical underpinning in the psychology literature by Moors and De Houwer.$^{(}$\!\citep{moors/dehouwer:06-pb}$^{\!)}$
Imperfect and fallible recognition, categorization, and procedural responses have been widely documented,%
$^{(}$\!\citep{chou/etal:90-ee,kaivanto/etal:14-eb,goldstein/taleb:07-jpm,vanlehn:90,frederick:05-jep}$^{\!)}$ 
for which general theoretical underpinning may be obtained from e.g. Philippe Jehiel's concept of analogy-based expectations equilibrium.$^{(}$\!\citep{jehiel:05-jet}$^{\!)}$}

\begin{table}[h]
\centering
\caption{Email recipients' coexisting choice criteria.}
\label{tab:ccc}
\begin{tabular}{lp{14.4cm}}
\hline\hline
$c\!=\!1$\T & 
Normative deliberation: characterized by the internal-consistency axioms of completeness, transitivity, independence of irrelevant alternatives (iia), continuity, Bayesian updating, and time consistency (i.e. exponential discounting). \B\\
$c\!=\!2$ & 
Behavioral: characterized by the weakening of iia, Bayesian updating, and time consistency (i.e. to hyperbolic discounting), as per the behavioral decision making literature.\B\\
$c\!=\!3$ & 
Impulsively click through: characterized by dominance of visceral emotions, which suppress and displace deliberative reasoning; the remaining consistency axioms are abandoned. \B\\
$c\!=\!4$ & 
Routinely click straight through: characterized by routinization and automaticity; again, the remaining consistency axioms are abandoned. \B\\
\hline
\end{tabular}
\end{table}

In general the choice-criterion selection probability will be conditional upon the decision maker's State of Mind, which in turn depends on an array of subject- and task-specific variables.
The net effect of all such variables determines an individual's probability of adopting a given choice criterion $c$ at a given point in time, which we denote by $\pi^c_{it}$. Note that we necessarily have $0\leq \pi^c_{\text{it}} \leq 1$ and $\sum_{c=1}^{C}\pi^c_{\text{it}}=1$ for all individuals $i$ and time-points $t$.

Figure \ref{fig:stripped-down-SoM} illustrates a single agent's stochastic State-of-Mind response to an arbitrary email. 
This begins with the diamond-within-a-circle chance node, whereby the incoming email probabilistically triggers one of the four State-of-Mind choice criteria. 
The fact that the `Routine' ($c\!=\!4$) and `Impulsive' ($c\!=\!3$) choice criteria override the possibility of sufficient deliberation to result in a `quarantine' choice with probability $\rho=1$ is indicated by the absence of these respective edges. The email recipient's incomplete information -- over whether the email is benign or malicious -- is reflected in the broken-line information sets surrounding terminal-node payoffs. 

\makeatletter
\newdimen\tempa
\newdimen\tempb
\pgfdeclareshape{diamond in circle}{
\inheritsavedanchors[from=diamond] 
\inheritsavedanchors[from=circle] 
\inheritanchorborder[from=circle]
\inheritanchor[from=circle]{center}
\inheritanchor[from=circle]{radius}
\inheritanchor[from=circle]{north}
\inheritanchor[from=circle]{south}
\inheritanchor[from=circle]{east}
\inheritanchor[from=circle]{west}
\inheritanchor[from=circle]{anchorborder}
  \saveddimen\radius{%
    %
    %
    \pgf@ya=.5\ht\pgfnodeparttextbox%
    \advance\pgf@ya by.5\dp\pgfnodeparttextbox%
    \pgfmathsetlength\pgf@yb{\pgfkeysvalueof{/pgf/inner ysep}}%
    \advance\pgf@ya by\pgf@yb%
    %
    %
    \pgf@xa=.5\wd\pgfnodeparttextbox%
    \pgfmathsetlength\pgf@xb{\pgfkeysvalueof{/pgf/inner xsep}}%
    \advance\pgf@xa by\pgf@xb%
    %
    %
    \pgf@process{\pgfpointnormalised{\pgfqpoint{\pgf@xa}{\pgf@ya}}}%
    \ifdim\pgf@x>\pgf@y%
        \c@pgf@counta=\pgf@x%
        \ifnum\c@pgf@counta=0\relax%
        \else%
          \divide\c@pgf@counta by 255\relax%
          \pgf@xa=16\pgf@xa\relax%
          \divide\pgf@xa by\c@pgf@counta%
          \pgf@xa=16\pgf@xa\relax%
        \fi%
      \else%
        \c@pgf@counta=\pgf@y%
        \ifnum\c@pgf@counta=0\relax%
        \else%
          \divide\c@pgf@counta by 255\relax%
          \pgf@ya=16\pgf@ya\relax%
          \divide\pgf@ya by\c@pgf@counta%
          \pgf@xa=16\pgf@ya\relax%
        \fi%
    \fi%
    \pgf@x=\pgf@xa%
    %
    %
    \pgfmathsetlength{\pgf@xb}{\pgfkeysvalueof{/pgf/minimum width}}%
    \pgfmathsetlength{\pgf@yb}{\pgfkeysvalueof{/pgf/minimum height}}%
    \ifdim\pgf@x<.5\pgf@xb%
        \pgf@x=.5\pgf@xb%
    \fi%
    \ifdim\pgf@x<.5\pgf@yb%
        \pgf@x=.5\pgf@yb%
    \fi%
    %
    %
    \pgfmathsetlength{\pgf@xb}{\pgfkeysvalueof{/pgf/outer xsep}}%
    \pgfmathsetlength{\pgf@yb}{\pgfkeysvalueof{/pgf/outer ysep}}%
    \ifdim\pgf@xb<\pgf@yb%
      \advance\pgf@x by\pgf@yb%
    \else%
      \advance\pgf@x by\pgf@xb%
    \fi%
  }
\backgroundpath{
    \tempa=\radius
    \pgfmathsetlength{\pgf@xb}{\pgfkeysvalueof{/pgf/outer xsep}}%
    \pgfmathsetlength{\pgf@yb}{\pgfkeysvalueof{/pgf/outer ysep}}%
    \ifdim\pgf@xb<\pgf@yb%
      \advance\tempa by-\pgf@yb%
    \else%
      \advance\tempa by-\pgf@xb%
    \fi%
    \pgfpathmoveto{\centerpoint\advance\pgf@x by\radius}%
    \pgfpathlineto{\centerpoint\advance\pgf@y by\radius}%
    \pgfpathlineto{\centerpoint\advance\pgf@x by-\radius}%
    \pgfpathlineto{\centerpoint\advance\pgf@y by-\radius}%
    \pgfpathclose%
  }
\behindbackgroundpath{
    \tempa=\radius%
    \pgfmathsetlength{\pgf@xb}{\pgfkeysvalueof{/pgf/outer xsep}}%
    \pgfmathsetlength{\pgf@yb}{\pgfkeysvalueof{/pgf/outer ysep}}%
    \ifdim\pgf@xb<\pgf@yb%
      \advance\tempa by-\pgf@yb%
    \else%
      \advance\tempa by-\pgf@xb%
    \fi%
    \pgfpathcircle{\centerpoint}{\tempa}%
  }
}
\makeatother

\vspace{0.4cm}
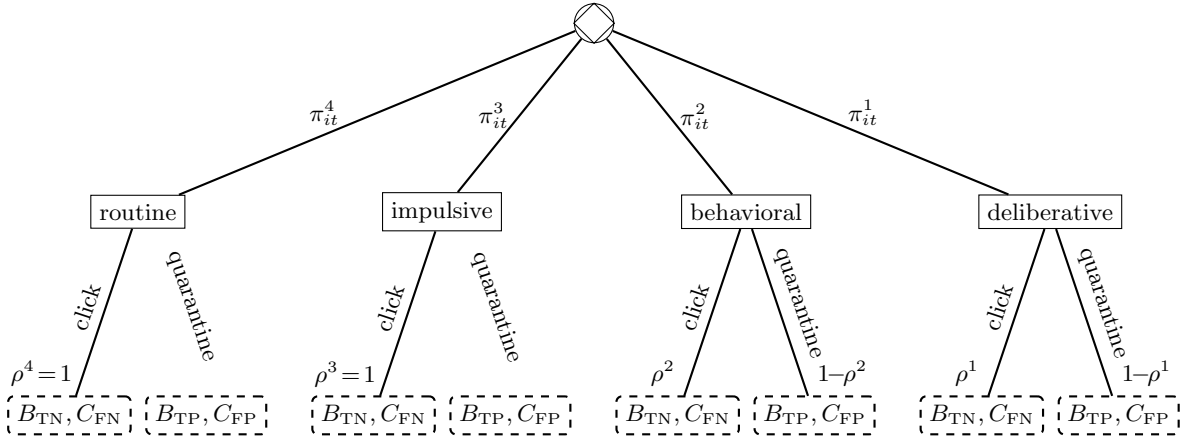
\begin{figure}[htbp]
	\centering
	\caption{An agent's stochastic State-of-Mind response to an email.}
	\label{fig:stripped-down-SoM}
\begin{tikzpicture}[font=\footnotesize,edge from parent/.style={draw,thick}]
    \tikzstyle{solid sq node}=[rectangle,draw,inner sep=2.2,fill=black];
    \tikzstyle{hollow sq node}=[rectangle,draw,inner sep=3.2];
    \tikzstyle{solid node}=[circle,draw,inner sep=1,fill=black];
    \tikzstyle{hollow node}=[diamond in circle,draw,inner sep=5.2];
    \tikzstyle{info set node}=[draw, dashed, rounded corners=2pt, inner sep=3.2];
    \tikzstyle{level 1}=[level distance=25mm,sibling distance=40mm]
    \tikzstyle{level 2}=[level distance=27mm,sibling distance=18mm]
  \node(0)[hollow node]{}
    child{node[hollow sq node]{routine}
      child{node[info set node]{{$B_{\text{TN}},C_{\text{FN}}$}}edge from parent node[above,sloped]{click} node[left,xshift=-8,yshift=-23]{$\rho^4\!=\!1$}}
      child{node[info set node]{$B_{\text{TP}},C_{\text{FP}}$}edge from parent[draw=none] node[above,sloped]{quarantine}}
      edge from parent node[left,xshift=-10]{$\pi^4_{it}$}
    }
    child{node[hollow sq node]{impulsive}
      child{node[info set node]{{$B_{\text{TN}},C_{\text{FN}}$}}edge from parent node[above,sloped]{click} node[left,xshift=-8,yshift=-23]{$\rho^3\!=\!1$}}
      child{node[info set node]{$B_{\text{TP}},C_{\text{FP}}$}edge from parent[draw=none] node[above,sloped]{quarantine}}
      edge from parent node[left,xshift=0]{$\pi^3_{it}$}
    }
    child{node[hollow sq node]{behavioral}
      child{node[info set node]{{$B_{\text{TN}},C_{\text{FN}}$}}edge from parent node[above,sloped]{click} node[left,xshift=-10,yshift=-23]{$\rho^2$}}
      child{node[info set node]{$B_{\text{TP}},C_{\text{FP}}$}edge from parent node[above,sloped]{quarantine$\;$} node[right,xshift=10,yshift=-23]{$1\!\!-\!\!\rho^2$}}
      edge from parent node[right,xshift=0]{$\pi^2_{it}$}
    }
    child{node[hollow sq node]{deliberative}
      child{node[info set node]{{$B_{\text{TN}},C_{\text{FN}}$}}edge from parent node[above,sloped]{click} node[left,xshift=-10,yshift=-23]{$\rho^1$}}
      child{node[info set node]{$B_{\text{TP}},C_{\text{FP}}$}edge from parent node[above,sloped]{quarantine$\;$} node[right,xshift=10,yshift=-23]{$1\!\!-\!\!\rho^1$}}
      edge from parent node[right,xshift=10]{$\pi^1_{it}$}
	    };
	  \end{tikzpicture}
	  \\ \vspace{1em} 
	  \scriptsize{Note: Ex ante the agent is uncertain about an email's true nature. The payoff at each terminal node is therefore either a benefit due to correct classification (True Positive or True Negative), or a cost due to incorrect classification (FP or FN).  
	  }
	  \end{figure}

The email recipient is one of many agents who interact in a strategic phishing game. We analyze an attacker's optimal response to Figure \ref{fig:stripped-down-SoM} in Section \ref{sec:stepping-stone-penetration}, and we discuss the model's implications for organizational security policy in Section \ref{sec:security_policy}. Before doing so, we complete the model by expanding the $\pi^c_{it}$ expression for an agent $i$ at time $t$. In general, $\pi_{it}^c$ is operationalized through a probability distribution that may be conditional upon: the characteristics of the decision maker $X_{it}$, the situational context $Z_{it}$,  and  the  attributes of the present choice task $\boldsymbol{\alpha}_{t}$.

\begin{subequations}
\begin{align}
\label{eq:general-choice-criterion-probabilities}
\pi^c_{it} &= \pi^c\big(X_{it}, Z_{it}, {\boldsymbol{\alpha}}_{t}\big),
\qquad 0\leq \pi^c_{it} \leq 1, 
\qquad \sum_{c=1}^{C}\pi^c_{it}=1 ,\\
\label{eq:characteristics-as-a-function}
X_{it}&=f\Big(\Gamma_i,\;\left\{Z_i\right\}_{<t},\;\left\{{\boldsymbol{\alpha}}\right\}_{<t},\;\left\{D_i\right\}_{<t}\Big)\;\; .
\end{align}
\end{subequations}
The current characteristics $X_{it}$ of agent $i$ are jointly determined by their stable psychological traits $\Gamma_i$, and by the history of: decision contexts $\{Z_i\}_{<t}$, decision-attributes $\{{\boldsymbol{\alpha}}\}_{<t}$, and decision-outcomes $\{D_i\}_{<t}$ that constitutes their current set of experiences. 

In order to develop a tractable expression for $\pi_{it}^c$ we generalize the notion of match quality introduced in the SDT literature$^{(}$\!\citep{kaivanto:14-ra}$^{\!)}$ and we specialize the vectors appearing in  \eqref{eq:general-choice-criterion-probabilities} to the phishing-email application. 
For this application, the context $Z_{it}$ is that in which the agent receives his emails. 
An agent whose context $Z_{it}$ and recent context history $\{Z_i\}_{<t}$ leaves him stressed, distracted, or hungry, will be less likely to respond deliberatively. 
The implications of this observation for personal practice and organizational security policy are clear,\footnote{These are discussed further in Section \ref{sec:security_policy},} and so we suppress $Z_{it}$ hereafter to focus on the strategic interaction between attackers and recipients. 
For simplicity we also suppress time subscripts hereafter to focus on the short-run implications of the model.

Let us consider a phishing email with attributes ${\boldsymbol{\alpha}}$ constructed within a finite attribute space $\mathcal{A}=[0,1]^A$. Each of the $A$ components of email ${\boldsymbol{\alpha}}$ captures the emphasis that it places on each of $A$ possible cues. 
The attacker chooses which cues to emphasize in order to influence the recipient's State of Mind. 
This determination of email `content' is the attacker's primary decision variable. 

The attacker is nevertheless constrained, in that increasing the emphasis placed on any one cue necessarily diminishes the emphasis  on the others.
We model this constraint by requiring $||{\boldsymbol{\alpha}}||\leq1$.

The salient characteristics of the recipient are his idiosyncratic susceptibility to each type of cue $S_i$, and his baseline propensity $\chi^c_i$ to apply each choice criterion $c$.\footnote{The baseline propensity to adopt a deliberative choice criterion $\chi^1_i$ is a stable trait$^{(}$\!\citep{parker/etal:17-bdm}$^{\!)}$ that can be measured by the cognitive Reflection Test of Frederick$^{(}$\!\citep{frederick:05-jep}$^{\!)}$, or the Decision-Making Competence scale of Parker and Fischhoff$^{(}$\!\citep{parker/fischhoffetal:05-bdm}$^{\!)}$.} $S_i$ is an $C\times A$ dimensional matrix, each row of which $\boldsymbol{s}^c_i$ specifies the effectiveness of each possible cue type in invoking the choice criterion $c$. The agent's characteristics $X_i$ are therefore a matrix in $[0,1]^{A\times C}\times(\mathbb{R}^+)^C$, each row of which is a pair $\{\boldsymbol{s}_i^c,\chi_i^c\}$ that will determine the match quality between the attacker's choice of email cues ${\boldsymbol{\alpha}}$, and the susceptabilities of the receiving agent $i$. 

We may now extend the approach of Kaivanto$^{(}$\!\citep{kaivanto:14-ra}$^{\!)}$ by defining the choice-criterion-specific match-quality function   $m^c\,:\,[0,1]^A\times[0,1]^A\times\mathbb{R}^+\to\mathbb{R}^+$, such that
\begin{equation}
    m^c_{i}({\boldsymbol{\alpha}}) = m^c({\boldsymbol{\alpha}},\boldsymbol{s}^c_i,\chi^c_{i})\qquad \forall\,c\in \mathcal{C}\;\;.
\end{equation}
For illustrative purposes, the simplest non-degenerate functional form for $m^c$ would be the separable linear specification 
\begin{equation}
    m^c_{i}({\boldsymbol{\alpha}}) = \chi^c_{i}+\boldsymbol{s}^c_i\cdot{\boldsymbol{\alpha}}\;\;,
\end{equation}
where $\cdot$ denotes the vector dot product. 

Agent $i$'s choice-criterion-selection probabilies for a given email with cue bundle ${\boldsymbol{\alpha}}$ may then be defined in terms of the match-quality functions as follows: 
\begin{equation}
    \pi_i^c({\boldsymbol{\alpha}})=\frac{m^c_i({\boldsymbol{\alpha}})}{\sum_{c\in \mathcal{C}}m^c_i({\boldsymbol{\alpha}})}\qquad \forall\,c\in \mathcal{C}\;\;.
\end{equation}

\subsection{Contrast with normatively rational deliberative special case}
\label{sec:contrast}
Under a normative decision-theoretic model of email-recipient decision making it is difficult to explain the existence of phishing as an empirical phenomenon. 
Normatively rational decision making is a special case of the coexisting-choice-criteria model in which $\pi^1=1$ and $\pi^2=\pi^3=\pi^4=0$. 
If all email recipients were characterized by choice-criterion \#1 alone, then the success of an email phishing campaign would be determined entirely by factors largely outside the attacker's control: the benefit from correctly opening a non-malicious email ($B_{\text{TN}}$), the cost of erroneously quarantining non-malicious email ($C_{\text{FP}}$), the cost of erroneously opening a malicious email ($C_{\text{FN}}$), and the benefit of correctly quarantining a malicious email ($B_{\text{TP}}$). 
Instead, variation in phishing campaigns' success rate is driven by factors that do not directly affect $B_{\text{TN}}, C_{\text{FP}}, C_{\text{FN}}$ and $B_{\text{TP}}$.$^{(}$\!\citep{rusch:99,mitnick/simon:02,hadnagy:11,oliveira/etal:17}$^{\!)}$ 

It is straightforward to explain the existence of phishing and its empirical characteristics under a coexisting-choice-criteria model of email-recipient behavior in which\; $\pi^1<1$\; and\; $\pi^2,\pi^3,\pi^4>0$. 
For instance choice-criterion \#4 (routine, automaticity) is triggered by a phishing email that masquerades as being part of the normal work flow by exploiting rich contextual information about the employee, the organizational structure (e.g. boss' and colleagues' names, responsibities, and working practices), and current organizational events and processes. 
Here the email recipient simply does not engage in a deliberative process to evaluate whether the email should be opened or not. 

In contrast, phishing ploys designed to trigger choice criterion \#3 (impulsively click through) employ what Robert Cialdini calls the \emph{psychological principles of influence} (see Section \ref{sec:human-element}).$^{(}$\!\citep{cialdini:07,rusch:99,mitnick/simon:02,hadnagy:11,oliveira/etal:17}$^{\!)}$ 
Importantly, there is variation between individuals in their susceptibility to particular levers of psychological influence.$^{(}$\!\citep{oliveira/etal:17,vishwanath/etal:11-dss,williams/etal:17-chb}$^{\!)}$ 
For instance scarcity\footnote{e.g. Don't miss out on this `once-in-a-lifetime opportunity!'} and authority\footnote{e.g. law enforcement officers, tax officials} have been found to be more effective for young users, while reciprocation\footnote{the tendency to repay in kind even though there is no implicit obligation to do so} and liking/affinity\footnote{the tendency to comply with requests made by people whom the user likes or with whom the user shares common interests or common affiliations} have been found to be more effective for older users.$^{(}$\!\citep{oliveira/etal:17}$^{\!)}$ 
These observations motivate the agent-specific subscript $i$ in $\pi^c_i$ and $m^c_i$, and they are important in establishing the constrained-optimal APT attack pattern in the following Subsection.

None of the aforementioned psychological levers would be effective if email users were solely $c\!\equiv \!1$ normatively rational deliberators. Similarly, the well-documented effects of commitment,\footnote{Also referred to as `consistency'. People feel obliged to behave in line with -- consistently with -- their previous actions and commitments.} 
perceptual contrast,\footnote{Making an option seem attractive by framing it with respect to an option that is 
(contrived to be) noticeably less attractive.}
and social proof \footnote{People conform with majority social opinion, even when this manifestly contradicts immediate personal perception, as in e.g. the Stanford Prison Experiment.} (see$^{(}$\!\citep{rusch:99,mitnick/simon:02,hadnagy:11,oliveira/etal:17}$^{\!)}$) are naturally explained by the existence of coexisting choice criteria.
 
\subsection{Stepping-stone penetration}
\label{sec:stepping-stone-penetration}
Forensic investigations of APT attacks have found that the initial breach point is typically several steps removed from the ultimate information-resource target(s). 
Deliberation-based models of normatively rational decision making offer no particular insight into this empirical regularity. 
In contrast, the coexisting-choice-criteria model encodes differentiation with which the stepping-stone penetration pattern may be recovered as a constrained-optimal attack vector. 

Let us consider an attacker who wishes to achieve a click-through from one of a minority subset of $m$ target individuals within an organization consisting of $n$ members. The target individuals may be those who can authorize expenditure, or those with particular (e.g. database) access rights. 
The attacker's strategy at any given point in time consists of a choice of cue-bundle ${\boldsymbol{\alpha}}_k$, taken to solve the program
\begin{equation}\label{attackers_objective}
    \max_{{\boldsymbol{\alpha}}_k} \;\;\;\sum_{i=1}^m\sum_{c=1}^C\,\pi_i^c({\boldsymbol{\alpha}}_k)\cdot\rho_i^c\cdot V \;\; - \;e({\boldsymbol{\alpha}}_k) \qquad \text{s.t.}\qquad ||{\boldsymbol{\alpha}}_k||\leq 1\;\;,
\end{equation}
where $\pi^c_i({\boldsymbol{\alpha}}_k)$ is the probability with which an individual will adopt choice criterion $c$ given the cues present in phishing email ${\boldsymbol{\alpha}}_k$, where $\rho^c$ is the probability of click-through given choice criterion $c$, where $V$ is the expected value of a successful attack, and where $e({\boldsymbol{\alpha}}_k)$ is the cost of the effort expended in the production and distribution of email ${\boldsymbol{\alpha}}_k$. 
This formulation accords with the near-zero marginal cost of including additional recipients to any existing email.$^{(}$\!\citep{shapiro/varian:98,anderson:08}$^{\!)}$ 

The attacker may send one, or more, emails ${\boldsymbol{\alpha}}_k$. 
Each email may be designed to induce one particular State-of-Mind $c$, or could in principle adopt a mixed strategy. 
However, since (by construction and by necessity) $\sum_{c\in\mathcal{C}}\pi^c_i=1$, any mixture of asymmetrically effective pure strategies must be strictly less effective than at least one pure strategy. 
We therefore proceed by characterizing the available pure strategies on the basis of the phishing literature,$^{(\!}$\citep{rusch:99,mitnick/simon:02,hadnagy:11}$^{\!)}$ before eliminating strictly dominated strategies.

\vspace{0.3cm}
\begin{table}[h]
\caption{Choice-criterion targeting characteristics.}
\label{tab:c_characteristics}
\vspace{-0.5cm}
\begin{center}
\begin{minipage}[t]{5.5in}
\begin{tabular}{lcccc}
\hline\hline
Choice criterion&Effort&Click-through prob.~~&\multicolumn{2}{c}{{Selection prob.%
\!\footnote{\,i.e. $\displaystyle\max_{\boldsymbol{\alpha}}\{\pi^c\}$}
}}\T\B\\
\cline{4-5}
\multicolumn{1}{c}{$c$~~~~~}&$e\big(\underset{{\boldsymbol{\alpha}}}{\argmax}\{\pi^c\}\big)$&$\rho^c$&\T Prior&\T\BB Posterior  \\
\hline
$c\!=\!1$: Deliberative \T & low & negligible&high &high \\ 
$c\!=\!2$: Behavioral& low & low& med & med\\ 
$c\!=\!3$: Impulsive & low & 1& low & low \\
$c\!=\!4$: Routine \B& high & 1& low & high\\
\hline 
\end{tabular}
\end{minipage}
\end{center}
\end{table}
\vspace{-0.5cm}

The quantities summarized in Table \ref{tab:c_characteristics} determine the costs and expected benefits to the attackers of targeting choice criterion $c$ through their choice of ${\boldsymbol{\alpha}}$. There are two values of the selection probability $\displaystyle\max_{\boldsymbol{\alpha}}\{\pi^c\}$ for each choice criterion $c$: the prior likelihood of invoking that criterion, without insider information, and the posterior likelihood once access to such insider information is obtained. Insider information does not affect the attackers' ability to invoke choice criteria $c\in\{1,2,3\}$, but it does greatly aid the attacker's ability to `spoof' (i.e. simulate) a routine email from a trusted colleague, and hence it substantially increases the posterior selection probability for $c=4$. 
The mechanism by which attackers may gain such insider information is the successful phishing of a non-target member of the organization.

The most immediate implication of Equation \eqref{attackers_objective} and Table \ref{tab:c_characteristics} is that the Deliberative strategy is strictly dominated by the Behavioral strategy, due to the negligible click-through probability of the former. 
We next observe that the Behavioral strategy is, in turn, strictly dominated by the Impulsive strategy whenever 
\begin{equation}\label{criterion1}
    \rho^2<\frac{\max_{\boldsymbol{\alpha}}\{\pi^3\}}{\max_{\boldsymbol{\alpha}}\{\pi^2\}}\;\;,
\end{equation}
That is whenever the expected click-through probability under a Behavioral choice criterion is less than the relative ease of invoking the Behavioral state compared to invoking the Impulsive state. Table \ref{tab:c_characteristics} suggests that this criterion is typically satisfied. 

Next we consider the case of an attacker who has no insider information. In this case it is trivial to see that an email which aims to invoke the Impulsive choice criterion strictly dominates an email which aims to invoke the Routine choice criterion, due to the lower effort cost of the former. The respective probabilities of successfully gaining a click-through from a target individual are then: 
\begin{equation}\nonumber
    \text{Prob.}\left(\substack{{\text{non-target}}\\{\text{clickthrough}}}\right) = 1\!\!-\!\!(1\!\!-\!\!\max_{\boldsymbol{\alpha}}\{\pi^3\})^{n\!-\!m} \quad>\quad 1\!\!-\!\!(1\!\!-\!\!\max_{\boldsymbol{\alpha}}\{\pi^3\})^m =  \text{Prob.}\left(\substack{{\text{target}}\\{\text{clickthrough}}}\right)
\end{equation}
which demonstrates that there is a greater likelihood of the attacker gaining a click-through from a non-target individual than from a target individual in any attack without insider information. 
Note that this conclusion would be further strengthened if we were to assume that target individuals were less susceptible to phishing attacks than the average individual. 

The attackers' first attempt therefore has three possible outcomes: 
(i) they may have successfully achieved their objective, 
(ii) they may have gained insider information by achieving a non-target click-through, or 
(iii) they may have achieved nothing. 
In the first case the attackers move on to acquire and exfiltrate the information. 
In the third case the situation is unchanged, and so the phishing campaign is continued with further broadcast of phishing email(s) containing (possibly modified) Impulsive cues. 
But in the second case insider information is obtained, whereby the posterior click-through likelihoods of Table \ref{tab:c_characteristics} become operative. 
In this case, it is evident from Table \ref{tab:c_characteristics} that an email which aims to invoke the Routine choice criterion is likely to dominate an email which aims to invoke the Impulsive criterion, specifically whenever 
\begin{equation}\label{criterion2}
    \frac{e\big(\text{argmax}_{\boldsymbol{\alpha}}\{\pi^4\}\big)}{e\big(\text{argmax}_{\boldsymbol{\alpha}}\{\pi^3\}\big)}<\frac{\max_{\boldsymbol{\alpha}}\{\pi^4\}}{\max_{\boldsymbol{\alpha}}\{\pi^3\}}\;\;.
\end{equation}
Thus the attacker's optimal approach is likely to lead to a `stepping-stone' attack, wherein a non-target individual is first compromised by invoking an impulsive choice criterion, so that a target individual can then be compromised by using insider information to invoke a Routine choice criterion. Sufficient conditions for this to be the most likely outcome are those of Table \ref{tab:c_characteristics} and inequalities \eqref{criterion1} and \eqref{criterion2}.

\subsection{Implications for Organizational Security Policy}
\label{sec:security_policy}
The model we present has important implications for organizational security policy. Let us first consider the cultural and procedural aspects of organizational security, before turning to specific implications for email security training and evaluation.

In Section \ref{sec:cccm} we noted the potential importance of the situational context $Z_{it}$ in which an email is received. 
For example, it is well-known that an individual who is under intense time-pressure is less likely, if not not simply unable, to engage in deliberative decision making.$^{(}$\!\citep{hwang:94-im,maule/edland:97,steigenberger/etal:17}$^{\!)}$ 
The present model makes plain the security-vulnerability dangers of highly routinized email-processing practices, even if these would otherwise be efficient. 
Relatedly, it is vital that organizational culture supports the precautionary verification of suspicious messages, since any criticism of such verification practices is likely to increase the risk of behavioral click-throughs in future. These observations suggest that Information Security Officers should actively engage with wider aspects of organizational culture and practices.

The model also yields specific procedural implications for email security training. 
It is clear that the direct effect of a training course in which participants consciously classify emails as either genuine or malicious would be to reduce $\rho^1$ (see Figure \ref{fig:stripped-down-SoM}), however for most individuals $\rho^1$ is already relatively low (see Table \ref{tab:c_characteristics}): given that an individual implements a deliberative choice criterion they are relatively unlikely to fall prey to a phishing attack. 
Section \ref{sec:stepping-stone-penetration} demonstrated that a strategic attacker would instead seek to exploit the much greater vulnerabilities of $\rho^3$ and $\rho^4$, and so training that focuses on reducing $\rho^1$ is likely to have limited effectiveness.

The challenge for Information Security Officers is that the vulnerabilities $\rho^3$ and $\rho^4$ are essentially fixed at 1.%
\footnote{In principle an organization could substantially reduce $\rho^4$ by implementing organization-wide 2-stage security procedures before any email link or attachment is opened, however such measures have not been widely adopted due to their efficiency cost (as discussed in Sections 1 and 2).} 
Once an Impulsive or Routine State of Mind takes over, click-through is a foregone conclusion. 
Training should therefore focus on reducing individuals' criterion-selection probabilities $\pi^3$ and $\pi^4$. 
There is evidence that an individual's propensity to act deliberatively can be raised through external interventions,$^{(}$\!\citep{frederick:05-jep}$^{\!)}$ and the coexisting-choice-criteria framework suggests that this could best be achieved by helping employees to understand: 
(i) their inherent vulnerability to phishing when making choices either Routinely or Impulsively, and 
(ii) the psychological ploys by which attackers may induce Impulsive or Routine States of Mind. 

Analogous implications exist for procedures which aim to test organizational security by means of simulated phishing emails. Where such a test is appended to a training module, it tests (at best) some combination of $\rho^1$ and $\rho^2$, because trainees will be aware that they are attempting to identify phishing emails. Furthermore, the literature on incentives suggests that where such a test is incentivized with some required pass-rate, it is likely to be less informative as to the true vulnerability level because it is more likely to generate a pure measure of $\rho^1$. 
Tests of security should therefore be blinded, for example by an unannounced simulation of an email attack. Moreover, such tests should be varied and repeated, since any single email $\boldsymbol{\alpha}$ can only contain one specific cue bundle, and so can only test an individual's susceptibility $\pi^c(\boldsymbol{\alpha})$ to that particular cue bundle.

\section{CONCLUSION}
As the basis for understanding and modeling the behavior of phishing targets, normative deliberative rationality proves wholly inadequate. 
This paper introduces a coexisting-choice-criteria model of decision making that generalizes both normative and `dual process' theories of decision making. 
We show that this model offers a tractable working framework within which to develop an understanding of phishing-email response behavior. 
This offers an improvement over existing SDT-based models of phishing-response behavior,$^{(}$\!\citep{kaivanto:14-ra,canfield/fischhoff:17-ra}$^{\!)}$ insofar as it avoids the commingling of peripheral-route-persuasion pathways. 

We also show that the framework may be usefully deployed in modeling the choices and tradeoffs confronted by APT attackers, who must make decisions about the nature, composition, and roll-out of phishing campaigns. 
We illustrate this by tackling a problem that has confounded conventional normative-rationality-based modeling approaches: Why do so many APT attacks follow a `stepping-stone' penetration pattern? 
Under the coexisting-choice-criteria model, the attacker faces a tradeoff between 
(i) designing an email that is highly targeted, invokes the `Routine' choice criterion, but requires detailed inside information, and
(ii) designing an email that cannot be targeted as effectively, invokes the `Impulsive' choice criterion, and requires only public information. 
However, success with (ii) provides the attacker with access to the inside information with which to implement (i). 
Thus the stepping-stone attack vector arises out of the attacker's tradeoffs precisely when confronting email users whose behavior is captured by the coexisting-choice-criteria model. 

We further demonstrate that the model provides new insights with practical relevance for Information Security Officers. 
We derive specific recommendations for information training and testing as well as for organizational procedures, practices, and policies. 
In particular, the model highlights the importance of considering the composite between the probability of being induced into State of Mind $c$ and the probability of then clicking through \emph{given} this State of Mind. 
Hence training must address the different State-of-Mind selection probabilities $\pi^c$ as well as the associated conditional click-through probabilities $\rho^c$.
Similarly, training-effectiveness testing -- assurance of learning, in effect -- must also cover the range of different State-of-Mind choice criteria. 
In light of the coexisting-choice-criteria model, the single-test-email approach should be deprecated. 

Finally, the coexisting-choice-criteria model highlights organizations' vulnerability to spear-phishing attacks that invoke automatic email processing routines. 
Working practices in most commercial, voluntary, and public-sector organizations presume that links and email attachments are benign when sent from within the organization or by customers, suppliers, or partner organizations. 
This is a major vulnerability that is as much a reflection of organizational culture as it is a reflection of explicit security protocols (or absence thereof). 
This suggests that Information Security Officers could -- and perhaps should -- be afforded a broader role in shaping organizational culture.



\newpage


\end{document}